\documentclass[aps,pra,preprint,titlepage,groupaddress,amssymb,amsmath,showpacs]{revtex4}
\usepackage{color}
\usepackage{graphicx}
\usepackage{graphics}
\usepackage{amsmath}

\begin{document}

\title{Automated Error Correction For Generalized Bell States}

\author{Pratyush Pandey}
\affiliation{Department of Electrical Engineering, Indian Institute of Technology Kanpur, Kanpur- 208016, India}

\author{Sriram Prasath E.}
\affiliation{Maulana Azad National Institute of Technology, Bhopal - 462051, India}

\author{Manu Gupta}

\author{Prasanta K. Panigrahi}
\affiliation{Indian Institute of Science Education and Research - Kolkata,
Mohanpur, P.O. BCKV Campus Main Office, Mohanpur - 741252, India}

\begin{abstract}
A procedure is developed to automatically correct the bit flip and the arbitrary phase change errors in Generalized Bell states (GBS). The phase and parity characteristics of the GBS are encoded in ancilla bits without altering the state under consideration. The same information is then used to correct the state at a different point, through appropriate unitrary control operations. It is also shown that the distributed/indirect measurements on ancilla can yield the error syndrome. The restricted error correction circuit obtained for N-qubit entangled states of the Bell type, are generalized for the corresponding higher dimensions, thus opening up the possibility of designing of task specific error correction circuits.
\end{abstract}


\maketitle

\section{Introduction}
Entangled states play a key role in quantum computation \cite{Niel, Stol}, distributed quantum computing \cite{Sefa} and a plethora of quantum communication protocols \cite{Duan, Brie}. Being part of a quantum network naturally exposes them to the decohering effects of the environment \cite{Conn, Gran}, which may alter both the bits, as well as the relative phases between the superposed states. Recently, it has been shown that the entanglement characteristics of Genralized Bell States (GBS) can be ascertained without destroying them \cite{Pani, Gupt}, which has also been experimentally realized \cite{Jhar}. In this scheme, measurements were carried out on ancilla channels, connected with the entangled state, for determining both phase and parity of the state under investigation. This is significant since the state under consideration may be part of a larger network, where a specific quantum task is in progress. This immediately suggests the possibility of correcting errors in such an entangled state by comparing the input and output states and correcting the latter by suitable control operations.

This approach, although similar in spirit to the ones where a weak control pulse derived from the original pulse is utilized for rectifying the stronger pulse at a later point in time \cite{Kevi}, differs from the known error correcting algorithms. In the well studied quantum error correction codes, e. g., CSS codes \cite{Shor, Ste1, Ste2, Cald, Eker, Cale, Smol, Flet}, involving finite dimensional Hilbert spaces, a given state (may be unknown) is embedded in a suitably enlarged Hilbert space, changing the state under consideration to a different basis altogether. The errors in such states are detected typically through majority voting and corrected by suitable unitary operations, usually involving measurements. In contrast, the present scheme deals with maximally entangled states and encodes the nature of the state in additional ancillas. This encoded information is then used to correct the entangled state whenever required. Moreover, unlike the CSS codes, this can be extended to the case of $d$ dimensional states as well. A number of methods exist for correcting errors in quantum network \cite{Delgado1}, some of which are also self-correcting codes \cite{Delgado2}. A procedure for error correction, which does away with the unitary operations, has also been shown \cite{Braunstein}.

The algorithm developed with the aim of automated error correction is illustrated here for known (having the nature of state encoded on ancillas) multipartite maximally entangled states. By automated, it is meant that the algorithm is implemented by using unitary operators and without any measurements in between, thereby making it possible to implement it in a circuit without any outside involvement. Thus the automated error correction is synonymous to measurement less error correction circuit. This procedure is useful when we are interested in obtaining the corrected state in the output and we are not interested in the specific error syndrome. To locate a specific error syndrome, one needs to carry out a suitable measurement. Our primary interest is to develop an automated error correcting circuit for correction of a restricted class of errors in a restricted set of quantum states of particular importance. This set contains bit flip and arbitrary phase changes. The restricted set of states includes all maximally entangled, n-qubit generalization of Bell states, which are of the form,

\begin{equation}
|\psi_{x}^{\pm}\rangle=\frac{1}{\sqrt{2}}(|x\rangle\pm|\bar{x}\rangle)\mbox{.}
\end{equation}

Here, $x$ varies from 0 to $2^{n-1}-1$ and $\bar{x}=1^{\otimes n}\oplus x$ in modulo 2 arithmetic. It is easy to see that this state reduces to Bell and GHZ states for $n=2$ and $n=3$ respectively. These maximally entangled states are generalization of Bell states and thus we will be referring them as GBSs hereafter. They have wide applications in quantum computing and thus it is interesting to study possibility of designing error correction protocol for these states. 

As will be seen below, our procedure protects the GBS states from bit flip and arbitrary phase change errors, while being part of a quantum channel. Following symmetry/property of these states are the key points on which our algorithm hinges:
\begin{enumerate}
\item
A GBS state that undergoes phase flip or bit flip or both, will still only be a (albeit different) GBS state only.
\item
An arbitrary phase change (not a flip), will change the GBS to a non-maximally entangled state, which can be brought back to the set.  
\item
The fact that the GBSs can be non-destructively discriminated is used to compare the initial and final states, to check and correct the state under consideration.
\end{enumerate}

Hence the set of GBS forms a group under the restricted set of error operations considered here, taking into account the presence of ancilla in the intermediate steps. Using the above properties an error correction algorithm has been designed for correction of bit flip and arbitrary phase change errors. The algorithm first corrects the arbitrary phase shift (still a phase flip error may exist), thus converting the non-maximally entangled state to a GBS. Then it compares the phase and parity, of initial and final state of the GBS and in case they differ, appropriate corrective measures are taken. All the correction steps of this algorithm, use only control unitary operations and does not involve any projective measurements or outside involvements.

In Section 2, the method used for discrimination of GBS will be discussed, followed by the error correcting algorithm in Section 3. In Section 4, proof of the algorithm will be given. Finally in Section 5, the generalization of this algorithm to higher dimentions is shown, after which we conclude by illustrating the advantages of this method and directions of future work.

\section{Non-Destructive Discrimination Algorithm}

In principle, any set of orthogonal states can be discriminated in quantum mechanics, but LOCC \cite{walg, ghos} may not be sufficient, if the state is distributed among two or more players. Recently, an experimentally verified method for discriminating orthogonal entangled states have been demonstrated, which uses measurements along ancilla qubits \cite{Pani, Gupt, Jhar}. In case of a physically separated GBS, the ancilla interacts separately first with one qubit and is then quantum communicated to another system qubit, with which it interacts and so forth, for any other possible system qubits. Finally, the ancilla qubit is measured in a suitable basis, the output of which indicates the result of measuring the corresponding multiparticle observable on the qubits. Below, we briefly illustrate the procedure as depicted in part 1 of Fig{1} and make it explicit through the example of Bell States.

In the entangled state is represented by $|\psi_x\rangle$, and initial ancilla is represented by $|a_0\rangle$ for $|0\rangle$ or $|a_1\rangle$ for $|1\rangle$, based on initial state selected. Ancilla qubits, having results of phase determination, will be referred to as $|\phi\rangle$ and the parity determination qubits will be denotd as $|p_i\rangle, i \in (1,n-1)$. $H_a$ is the Hadamard operation on the ancilla qubit, while $Q(|x\rangle\rightarrow |y\rangle)$ means a C-NOT, operating with $|x\rangle$ as the control, and $|y\rangle$ as the target. $|{\psi_x}_j\rangle$ here represents the $j^{th}$ qubit of GBS. 

The ancilla $|\phi\rangle$, which yields the information about the relative phase of the GBS, is given by:
\begin{equation}
\begin{split}
|\psi_x\rangle |\phi\rangle = [(I_2^{\otimes n} \otimes H_a) \times (\bigotimes_{j=1}^{j=n}Q(|a_0\rangle \rightarrow |{\psi_x}_j\rangle)) \times(I_2^{\otimes n} \otimes H_a)] (|\psi_x\rangle|a_0\rangle)
\end{split}
\end{equation}

The $i^{th}$ parity qubit is given by:

\begin{equation}
|\psi_x\rangle|p_i\rangle = [ Q(|\psi_i\rangle \rightarrow |a_{0i}\rangle) \otimes Q(|{\psi_x}_{(i+1)}\rangle \rightarrow |a_{0i}\rangle)] (|\psi_x\rangle|a_{0i}\rangle)
\end{equation}

As an illustration, a two qubit Bell state has one parity and one phase qubit. The measurement results corresponding to various Bell states are given in the following table:
\[ \begin{array} {ccc}
\vspace{2mm}
\mbox{Bell State} & |p\rangle & |\phi\rangle \\
\vspace{2mm}
|\psi_1\rangle - \frac{1}{\sqrt{2}}(|00\rangle + |11\rangle) & |0\rangle & |0\rangle \\
\vspace{2mm}
|\psi_2\rangle - \frac{1}{\sqrt{2}}(|00\rangle - |11\rangle) & |0\rangle & |1\rangle \\
\vspace{2mm}
|\psi_3\rangle - \frac{1}{\sqrt{2}}(|01\rangle + |10\rangle) & |1\rangle & |0\rangle \\
\vspace{2mm}
|\psi_4\rangle - \frac{1}{\sqrt{2}}(|01\rangle - |10\rangle) & |1\rangle & |1\rangle \\
\end{array}\].

\section{Automated Error Correction Algorithm}

The algorithm is designed to correct any bit flip or arbitrary phase change errors in GBS, by using the non-destructive discrimination and comparing the initial and final states. The algorithm consists of two parts,  part 1 is the initial setup step to attain the phase and parity information of the GBS, without altering its state. Part 2 is the three step error correction algorithm, which can be executed repeatedly on demand, whenever it is required to restore the GBS to its initial state. Both the parts of the algorithm are clearly depicted in Fig{1}. 
\\
\\
Part 1: \\
Apply the discrimination algorithm to the initial error free GBS state and attain the parity [$p$] and phase [$\phi$] information of the states on the ancilla qubits.
\\
\\
Part 2: \\
When required to correct the error on this GBS, recreate the initial ancilla qubits based on the measurement results of part 1, and execute the following three steps; 

\begin{enumerate}
\item
This step removes the arbitrary phase shift error (still a phase flip may exist), through control unitary operations using an ancilla qubit, and thus converting the non-maximally entangled state to a GBS.

\item
Here the phase flip error is removed if it exists, by applying the phase discrimination algorithm and comparing the results with initial phase. In case a difference arises, the phase is corrected through control unitary operations

\item
This last step removes the bit flip errors if it exists, by applying the parity discrimination algorithm and comparing the results with initial parity. In case of a difference, the bit flip is corrected through control unitary operations

\end{enumerate}

We note that the process of measurement at the end of part 1 is not mandatory, instead a quantum channel can be used to communicate the information of the phase and parity of GBS and thus apply the correction algorithm when required. However, without measurement the correction algorithm can only be applied once on the GBS.

\begin{figure}
\begin{center}
\includegraphics[scale=0.45]{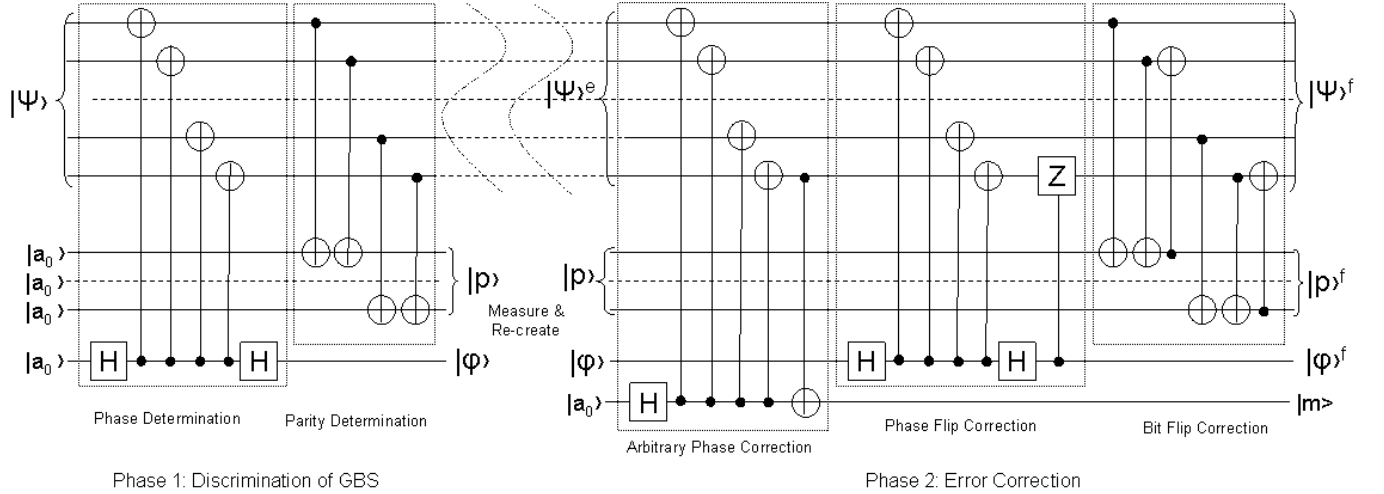}
\caption{\label{eca}Error Correction Algorithm.}
\end{center}
\end{figure}

The erroneous input GHZ state will be depicted with a superscript $'e(i)'$ and the final corrected state with superscript $\rho('f')$. All the final states of the phase and parity ancillas, will be denoted by a superscript $'f'$. Now, we explicate the three steps of the error correction algorithm.

\textbf{Step 1:} For correction of arbitrary phase error, take an ancilla qubit $|a_0\rangle$, and apply the control unitary operations as specified below,

\begin{equation}
\begin{split}
|{\psi_x}\rangle |m\rangle = [(I_2^{\otimes n} \otimes H_a) \times (\bigotimes_{j=1}^{j=n}Q(|a_0\rangle \rightarrow |{\psi_x}_j\rangle)) \times (Q|{\psi_x}_n\rangle \rightarrow |a_0\rangle) \times (I_2^{\otimes n} \otimes H_a)] (|{\psi_x}^e\rangle^{(i)} |a_0\rangle)
\end{split}
\end{equation}

After this operation, the GHZ state will be maximally entangled, with possible bit flip errors and other arbitrary phase errors will be passed to the ancilla. 

\textbf{Step 2:} For correcting the phase flip, one takes the phase qubit $|\phi\rangle$, which is initially determined in Eq.{2}. Once gain, we repeat the initial phase operation for the erroneous $|\psi_x\rangle$:

\begin{equation}
\begin{split}
|{\psi_x}^{e1}\rangle |\phi^f\rangle = [(I_2^{\otimes n} \otimes H_a) \times (\bigotimes_{j=1}^{j=n}Q(|\phi\rangle \rightarrow |{\psi_x}_j\rangle)) \times (I_2^{\otimes n} \otimes H_a)] (|{\psi_x}^{e1}\rangle|\phi\rangle)
\end{split}
\end{equation}

The phase qubit $|\phi^f\rangle$ attained is the comparison between the initial and the final phase of the state $|\psi\rangle$. If the two phases are different, $|\phi^f\rangle = |1\rangle$ else $|\phi^f\rangle = |0\rangle$.

To correct the phase of the state $|\psi\rangle$, we perform the operation given below, using the obtained $|\phi^f\rangle$,

\begin{equation}
|{\psi_x}^{e2}\rangle |\phi^f\rangle = [Q(|\phi^f\rangle \rightarrow |{\psi_x}_n\rangle \otimes Z)] (|{\psi_x}^{e1}\rangle|\phi^f\rangle)
\end{equation}

This equation indicates the operation of the Pauli matrix $Z$ on $x_n$ qubit of the state $|\psi\rangle$, if $|\phi^f\rangle = |1\rangle$, else no corrections are required.

\textbf{Step 3:} For correcting any bit flips in the state, we consider the $i^{th}$ parity qubit $|p_i\rangle$, where $p_i, i \in (1,n-1)$. Initially $|{p_i}\rangle$ is determined as per Eq.{3}. Now we repeat the initial parity operation for the erroneous $|\psi^{e2}\rangle$:

\begin{equation}
|{\psi_x}^{e2}\rangle|{p_i}^f\rangle = [Q(|{\psi_x}_i\rangle \rightarrow |p_i\rangle) \otimes Q(|{\psi_x}_{(i+1)}\rangle \rightarrow |p_i\rangle)] (|{\psi_x}^{e2}\rangle|p_i\rangle)
\end{equation}
parity qubit $|{p_i}^f\rangle$ attained is basically the comparison between the initial and the final parity, for $i^{th}$ and $(i+1)^{th}$ qubits, of the state $|\psi\rangle$. If the two parities are different, $|{p_i}^f\rangle = |1\rangle$ else $|{p_i}^f\rangle = |0\rangle$.

Finally, correct the state $|\psi\rangle$ using the parity qubit $|{p_i}^f\rangle$
\begin{equation}
|{\psi_x}^f\rangle|{p_i}^f\rangle = [Q(|{p_i}^f\rangle\rightarrow |{\psi_x}_{(i+1)}\rangle)] ({\psi_x}^{e2}\rangle|{p_i}^f\rangle)
\end{equation}
the operation flips the $|{\psi_x}_{(i+1)}\rangle$ qubit, if $|{p_i}^f\rangle = |1\rangle$. Now, repeat this procedure for $i \in (1,n-1)$. Note that all the operations used here are unitary, and hence can be physically implimented in circuits.

\section{Proof of the Algorithm}

Consider first the arbitrary phase correction, the circuit used is, to an extent similar to the phase determination circuit, however with no Hadamard and an extra controlled unitary Pauli-X operation in the end. This passes the arbitrary phase in state $|\psi\rangle$ to the ancilla qubit, converting it to a maximally entangled state. For illustration, consider the GBS with an arbitrary phase error:

\begin{eqnarray}
|\psi_{x}^{e1}\rangle |a_0\rangle & = &\frac{1}{\sqrt{2}}(|x\rangle \pm e^{i\delta}|\bar{x}\rangle)(|0\rangle)
\nonumber \\
& =&\frac{1}{\sqrt{2}}(|x\rangle \pm e^{i\delta} |\bar{x}\rangle)(|0\rangle + |1\rangle)
\nonumber \\
& =&\frac{1}{\sqrt{2}}[(|x\rangle \pm e^{i\delta} |\bar{x}\rangle)|0\rangle + (|\bar{x}\rangle \pm e^{i\delta} |x\rangle)|1\rangle]
 \end{eqnarray}
We now apply controlled unitary Pauli-X operation from $|{\psi_x}_n\rangle $ to $|a_0\rangle $ and re-arrange the terms as,
\begin{equation}
= \frac{1}{\sqrt{2}} (|x\rangle + |\bar{x}\rangle) ( |0\rangle \pm e^{i\delta_1} |1\rangle )
\end{equation}

For two qubit Bell States the above equation reduces to 

\begin{eqnarray}
 |\psi_{x}^{e1}\rangle &=& \frac{1}{\sqrt{2}}(|00\rangle + |11\rangle)(|0\rangle \pm e^{i\delta}|1\rangle)
\nonumber \\
&=& \frac{1}{\sqrt{2}}(|01\rangle + |10\rangle)(|1\rangle \pm e^{i\delta}|0\rangle)
\end{eqnarray}

Moving on to the phase flip correction, we can see that the operation 'controlled Pauli-Z' will selectively flip the phase of the state $|{\psi_x}^{e2}\rangle$, when the initial and the final phase of the state differs. This will cause the phase change to be corrected.

Now consider the parity checking. Let us assume that the first qubit transferred is correct, i.e., $|{{\psi_x}_1}^{e2}\rangle = |{{\psi_x}_1}\rangle$. In this case, look at the initial and final states of the parity qubit $|p_1\rangle$. If $|{{\psi_x}_2}^{e2}\rangle$ has not flipped, then $|p_1\rangle = |0\rangle$, and thus the qubit will remain unchanged.

Similarly, if $|{{\psi_x}_i}^{e2}\rangle$ is assumed to be undamaged, then $|{{\psi_x}_{(i+1)}}^{e2}\rangle$ will be flipped, only if it is not equal to $|{{\psi_x}_{i+1}}\rangle$. Thus, if $|{{\psi_x}_1}^{e2}\rangle$ is undamaged, by induction, all the other qubits that are damaged will be flipped, and therefore, the final qubit will be the same as the initial.

Now let us consider the case in which $|{{\psi_x}_1}^{e2}\rangle$ is damaged. That is, $|{{\psi_x}_1}^{e2}\rangle \ne |{{\psi_x}_1}\rangle$. Now, if $|{{\psi_x}_2}^{e2}\rangle$ is undamaged, the corresponding parity qubits will differ. This will cause a flip in $|{{\psi_x}_2}^{e2}\rangle$ which will finally make it $|\bar{{\psi_x}_2}\rangle$. If, on the other hand, $|{{\psi_x}_2}^{e2}\rangle = |\bar{{\psi_x}_2}\rangle$, then the parity qubits will match, and again, the final qubit will be the negation of the initial. Therefore, in any case, the final qubit will be $|\bar{{\psi_x}_2}\rangle$. This will go on for each $|{{\psi_x}_i}^{e2}\rangle$, so that finally each of them will be turned into $|\bar{{\psi_x}_i}\rangle$.

This means that, finally, $|{\psi_x}^f\rangle$ is equal to $|\psi_x\rangle$ or $|\Bar{\psi_x}\rangle$. However, based on the symmetry properity, it is known for GBS that, $|\psi_x\rangle$ and $|\Bar{\psi_x}\rangle$ are identical states.

\section{Generalization and Advantages}

Above automated error correction circuit can be generalized for qudits, based on the fact that the non-destructive discrimination algorithm, which is the backbone of our circuit, is easily extendable to higher dimensions \cite{Gupt}. Thus by extending our error correcting circuit in the same lines of discriminatory algorithm will result in the correction of errors in higher dimensions.  

To conclude, this algorithm corrects bit and arbitrary phase change errors on GBS. It needs to be emphasized that while correcting the errors, the entangled states retain their identity and hence can be part of an ongoing quantum operation, thus establishing the fact that there exists task specific error correction circuits, opens up the huge possibilities of study of such task specific error correction circuits which will correct a restricted set of error in a restricted set of quantum states of interest.

\section{Acknowledgment}

PP acknowledges useful discussions with Prof. V. Singh of HBCSE, where a part of this work was carried out. MG thanks Dr. A. Pathak for helpful discussions and for his valuable advice and support. We also thank Profs. S. L. Braunstein and M. A. Martin-Delgado for bringing relevant references to the authors' notice.

\end{document}